# Controlling spectral energies of all harmonics in programmable way using time-domain digital coding metasurface

Jie Zhao[1†], Xi Yang[2†], Jun Yan Dai[1†], Qiang Cheng[1,3]*, Xiang Li[2], Ning Hua Qi[1], Jun Chen Ke[1], Guo Dong Bai[1], Shuo Liu[1], Shi Jin[2]*, and Tie Jun Cui[1,3]*

[1]*State Key Laboratory of Millimeter Waves, Southeast University, Nanjing 210096, China*

[2]*National Mobil Communication Research Laboratory, Southeast University, Nanjing 210096, China*

[3]*Jiangsu Cyber-Space Science & Technology Co., Ltd., 12 Mozhou East Road, Nanjing 211111, China*

† These authors contributed equally to this work.

* Corresponding author: qiangcheng@seu.edu.cn; jinshi@seu.edu.cn; tjcui@seu.edu.cn

## Abstract

Modern wireless communication is one of the most important information technologies, but its system architecture has been unchanged for many years. Here, we propose a much simpler architecture for wireless communication systems based on metasurface. We firstly propose a time-domain digital coding metasurface to reach a simple but efficient method to manipulate spectral distributions of harmonics. Under dynamic modulations of phases on surface reflectivity, we could achieve accurate controls to different harmonics in a programmable way to reach many unusual functions like frequency cloaking and velocity illusion, owing to the temporal gradient introduced by digital signals encoded by '0' and '1' sequences. A theoretical model is presented and experimentally validated to reveal the nonlinear process. Based on the time-domain digital coding metasurface, we propose and realize a new wireless communication system in binary frequency-shift keying (BFSK) frame, which has much more simplified architecture than the traditional BFSK with excellent performance for real-time message transmission. The presented work, from new concept to new system, will find important applications in modern information technologies.

Nonlinear effects are typically observed during the interaction of electromagnetic (EM) waves and natural materials in a wide spectrum ranging from microwave to optical frequencies (*1-3*). Such phenomena are usually associated with the dielectric polarization responding in a nonlinear way to the intensity of incident field, resulting in radiation fields oscillating at new frequencies or propagating to new directions (*4-6*). Up to now, the nonlinear properties have been witnessed profound effects on applications in communication, optical storage, and all-optical computing (*7-9*). However, the mainstream nonlinearity generation methods suffer from various major problems. For example, a drawback of the conventional nonlinear crystal is inaccurate phase control of the dielectric polarizability, which hinders ability to perform elaborate wave manipulations for all harmonics (*10,11*). In addition, the frequency conversion requires sufficiently high optical intensities to enhance the nonlinear process (*12,13*), which may be limited by damage threshold of materials in practice.

Recent advances of metamaterials and metasurfaces have provided an unprecedented degree of freedom to manipulate the electromagnetic waves in subwavelength scale, giving rise to the enhanced nonlinearity with giant local fields, and thus offering new possibilities in controlling the intensity, phase, and polarization states of induced harmonics (*14-18*). Owing to the sensitivity of structural symmetry to the nonlinear susceptibilities, various shapes of meta-atoms have been employed in second-harmonic generation (SHG) by optimizing the element orientation as well as tuning the interaction between neighboring elements, making it possible to achieve the phase control over the local nonlinearity for the generated harmonics (*19-21*). Apart from the interesting SHG phenomena, nonlinear effects over higher-order harmonics have also received considerable attention and have been used in the third-harmonic generation (THG) and four-wave mixing (*22-25*). However, the efficiency of frequency conversion strongly depends on the input field intensity, forming a barrier of substantial nonlinear response to weak signals.

As an alternative approach, by using time-varying metasurface, it has been proved that the normal momentum component of the incident photon makes a difference on the two sides of the metasurface (*26,27*), enabling spatiotemporal modulation of electromagnetic waves and frequency shifting due to temporal phase gradient*28*. Nevertheless, this theoretical approach on one hand is lack of simple and elegant way to realize in application (*26-28*), on the other hand, it still remains a significant challenge to generate arbitrary wave manipulations for both linear and nonlinear frequency components with high precision and high efficiency.

Another route to achieve the nonlinear control lies in time-modulated antenna array (*29,30*). Through simple binary switching of diodes integrated into the array elements, one is able to tailor the radiation patterns at both central and harmonic frequencies, and thus benefit a plenty of applications in microwave engineering. However, this scheme lacks essential response to the external excitations as the natural materials since the harmonics are resulted from the Fourier components of the rapid switching over a built-in source, hereby fails to perceive and respond to external electromagnetic waves in a predefined way, limiting the application area of the method. Moreover, most of research efforts in this field are focused on amplitude modulation of the antenna element, since complicated circuit designs are usually required to implement the desired phase modulations over the high-order harmonics, which will inevitably increase the system complexity and the risk of performance degradation (*31*).

To solve the above difficulties, here we propose a new route to produce and control the nonlinearity of waves in free space with much stronger capabilities by exploiting a time-domain digital coding metasurface with dynamic reflection features. Inspired by spatial-domain digital coding metasurfaces (*32,33*), complex modulation strategies are employed to tailor the wave-matter interactions and frequency spectrum simultaneously, where discrete reflection phase states of metasurface are altered in digital coding sequences with a controlled manner. We demonstrate that nonlinear process will take place from the temporal modulation of incident waves on the metasurface with accurate controls of both amplitude and phase distributions for all harmonics. A rigorous theory is presented to describe the nonlinear process. To experimentally validate the theory, we fabricate a sample composed of periodic coding element loaded with varactor diode. Driven by different combinations of output voltages from an interface of field programmable gate array (FPGA), the single metasurface can operate many functions by controlling the time- domain digital coding states. This produces a time-domain programmable coding metasurface, which have powerful capabilities to generate and manipulate the nonlinearity, and could trigger some interesting phenomena such as frequency cloaking and velocity illusion. As an application of the time-domain digital coding metasurface, a binary frequency-shift keying (BFSK) communication system is explored within a novel framework in contrast to the classical heterodyne architectures. In the new BFSK system, the two basic carrier frequencies are synthesized directly via the metasurface without using complicated mixing processes, showing great advantages of the time-domain digital coding metasurface. The proposed theory and method will pave a way for simplified and compact communication and radar systems across the frequency range from acoustic, microwave to optics.

**Time-domain digital coding metasurface.** A coding metasurface is composed of digital units. For 1-bit coding, the two digital units '0' and '1' have either opposite phases or 0 and 1 transmission coefficients. However, the existing references (*32-35*) have only studied space-domain digital coding metasurfaces to control electromagnetic waves by using different spatial coding sequences. Here, we propose the concept of time-domain digital coding metasurface, in which the digital units are controlled by different periodic time sequences from FPGA, as shown in Fig. 1. We start to investigate interactions between the electromagnetic waves and the time-domain digital coding metasurface, which is composed of periodic coding elements loaded with varactor diodes. Then the reflectivity can be expressed as a periodic function of time and defined over one period as a linear combination of scaled and shifted pulses. Under the excitation of a monochromatic signal $E_i(f)$ with frequency $f_c$, the reflected signal from the time-domain digital coding metasurface can be derived as (see Supporting Online Material, SOM):

$$E_r(f) = a_0 E_i(f) + \sum_{k=1}^{\infty} [a_k E_i(f - kf_0) + a_{-k} E_i(f + kf_0)] \tag{1}$$

in which $f_0 = 1/T$ ($T$ is the period of the reflectivity function), and $a_k$ is the complex Fourier series coefficient at $kf_0$:

$$a_k = \frac{1}{M} Sa\left(\frac{k\pi}{M}\right) \exp\left(-j\frac{k\pi}{M}\right) \cdot \sum_{m=0}^{M-1} \Gamma_m \exp\left(-j\frac{2km\pi}{M}\right) = UF \cdot TF \tag{2}$$

in which, $M$ is the length of coding sequence in one period, $\Gamma_m$ is the reflectivity at the interval $(m\text{-}1)\tau < t < m\tau$, $\tau = T/M$ is the pulse width, and

$$TF = \sum_{m=0}^{M-1} \Gamma_m \exp\left(-j\frac{2km\pi}{M}\right), \quad UF = \frac{1}{M} Sa\left(\frac{k\pi}{M}\right) \exp\left(-j\frac{k\pi}{M}\right) \tag{3}$$

It is clear that $a_k$ can be regarded as the product of two terms: the time factor (TF) and unit factor (UF). The former is related to the modulation signal ($\Gamma_m$) within different time slots, while the latter is the Fourier series coefficient of the basic pulse with the pulse width $\tau$ and repeated with the period *T*. UF defines the basic spectrum property within one pulse, while TF represents the coding strategy.

It is obvious that the time-domain digital coding metasurface is able to generate a number of harmonics $f_c \pm kf_0$ by simply controlling the time-domain coding states around the central frequency. Meanwhile, it offers the possibility of nonlinear amplitude and phase manipulations for all harmonics by tuning the coefficients $a_k$ using an optimized modulation TF. We remark that the proposed method is essentially different from the traditional mixing techniques, and relies on the dynamic reflections on the coding metasurface to produce the spectral features of incident waves.

**Modulation methods.** To illustrate the nonlinear control capability to the spectral features of reflected waves by the time-domain digital coding metasurface quantitatively, we calculate the spectral distributions of all harmonics under the illumination of monochromatic plane wave. In fact, both amplitude and phase modulations of the reflectivity on metasurface could achieve efficient controls to the nonlinearity. SOM Figure S2 presents the results of amplitude modulation (AM) on the metasurface. An important feature of AM is the symmetry of amplitude spectra with respect to the central frequency, due to the fact that both +$k^{th}$ and -$k^{th}$ Fourier series components have equal amplitude. However, in some applications like communication systems, the message signals may not be dispersed simultaneously into the upper (+$k^{th}$) and lower (-$k^{th}$) harmonics, which leads to waste of energy and spectrum resources. On the other hand, the AM reflectivity is usually required less than unity in many situations, and hence part of incident energy has to be absorbed by the metasurface. This should be avoided in power-limited systems. More importantly, the zero$^{th}$-order harmonic (the central frequency) cannot be eliminated totally since the reflectivity is always positive in the whole period, which makes it impossible to realize frequency cloaking.

To circumvent the disadvantages, phase modulation (PM) of the time-domain reflectivity is explored to provide *asymmetric spectral responses* and promote the modulation efficiency. Figure 2 shows the spectral intensity distributions of all harmonics excited by the metasurface under different time coding sequences, in which the reflectivity is described as a signal with unity amplitude but digital phase states. Interestingly, we find that the zero$^{th}$-order harmonic is totally eliminated with a 1-bit coding sequence 010101… ($M$ = 2 and $T$ = 1μs), as the reflection phase changes between 0° and 180° periodically, as shown in Fig. 2A-B. This phenomenon can be ascribed to signal cancellation in each period from the anti-phase reflectivity, making the coefficient $a_0$ equal zero. This feature can also make the proposed metasurface be a ***frequency cloak***, since it transfers the incident energy at certain frequency to its higher-order harmonics completely, and the reflected and scattered energies at this frequency cannot be detected. Figure 2C-D illustrates another case with 2-bit coding 00-01-00-01-…, in which the signals of even-order harmonics (except $k$ = 0) are totally suppressed as the phase of the reflectivity switches between 0 °and 90 °periodically.

If more phase states are introduced in the PM signal, one can further remove unwanted harmonics and create ***asymmetric energy distributions*** in the whole spectra, as shown in Fig. 2E-H, where 2-bit coding 00-01-10-11-… and 11-10-01-00-… with anti-symmetric phase ramps are exploited to modulate the reflectivity with mirrored spectrum distributions. The reason for the observed behavior is that, when the reflectivity of the metasurface becomes complex and time-variant, the relationship $a_{-k} = a_k^*$ is no longer preserved in contrast to the AM scenario, resulting in the spectral asymmetry. It is also easy to find that the anti-symmetric phase ramps in Fig. 2E and 2G will lead to mirror transformation of spectrum, thereby translating the original signals from +$k^{th}$ harmonic to the corresponding -$k^{th}$ harmonic. The flexible controls of harmonic spectra can be used to create ***velocity illusions*** and new-architecture communication systems, which will be discussed in details in the experimental section.

**Experiments on nonlinear reflections.** For experimental demonstration, we design a PM time-domain digital coding metasurface, as illustrated in Fig. 3A, in which the zoomed-in view of the meta-atom is depicted in the inset. Two rectangular patches linked by a varactor diode (SMV-2019) are repeated periodically on the top of the substrate (F4B, $\epsilon_r = 2.65(1 - j0.001)$ and thickness=4mm). A number of narrow slots (width=0.15mm) are etched on the bottom plate for biasing the diodes through metallic via holes. To eliminate the electromagnetic leakage from slots, another ultrathin vinyl electrical tape (3M Temflex, thickness=0.13mm) backed by a metal layer is placed underneath the slots to prevent the wave penetration. Each element has the size of $18.8\times16.1\text{mm}^2$, and the geometric dimensions are optimized to achieve continuous adjustment of the reflection phase with varied biasing voltages near 3.6GHz. Commercial software, CST Microwave Studio, is used to quantify the spectral response of the metasurface when illuminated by a plane wave polarized in the $x$ direction. Figure 3B-C illustrates the simulated reflection amplitude and phase spectra as a function of the biasing voltage, showing high reflection amplitude and nearly 360° phase range in the spectra of interest, which is especially fit for PM in experiments.

A prototype composed of 16×16 meta-atoms is fabricated and measured in a microwave anechoic chamber (SOM Fig. S3). Figure 3D illustrates the measured harmonics distributions modulated by the 1-bit time-domain digital coding 01010101… with two phase states ($\varphi = 0°$ and $\varphi = 180°$) at different pulse durations, in which two voltages (0V and -9V) are chosen to bias the varactor diodes to reach '0' and '1' digital states. We clearly observe the harmonic generations using the time-domain digital coding metasurface as predicted. When the pulse duration $\tau$ increases from 0.8 to 6.4μs gradually, the harmonic energies are insensitive to the variance of $\tau$ since the amplitude of Fourier series is only associated with the coding sequence in modulation, as illustrated in SOM Fig. S4. Consistent to the theoretical calculations shown in Fig. 2B, the proposed metasurface transfers most of the incident energy at the central frequency to higher-order harmonics effectively, realizing good *frequency cloaking* performance. We remark that there is a small portion of the zero$^{\text{th}}$-order harmonic owing to the simplified modelling and measurement errors.

The frequency cloaking effect can be clearly observed from the far-field scattering pattern. Concerning the on-off dependency of the modulation signal with the binary coding sequence 01010101…, the measured scattering patterns of the metasurface in H-plane are presented in Fig. 3E. As the biasing circuit is switched on, the maximum scattering directivity for the central frequency ($0^{\text{th}}$-order harmonic) exhibits a significant drop by 15dB than that in the off state, corresponding to the good cloaking performance. The harmonics of the $\pm 1^{st}$ orders show nearly identical radiation patterns as a result of symmetric harmonic generations under the current coding sequence, as demonstrated in Fig. 3F.

For 2-bit coding sequences containing four digital states '00', '01', '10', and '11' with the corresponding biasing voltages 0, -6, -9, and -21V, we can achieve reflection responses with the same amplitude but discrete phase states 0°, 90°, 180°, and 270°, respectively. *The symmetry of the reflection spectra will be broken in the presence of time gradient.* This is confirmed by the measured results shown in Fig. 3G, in which the harmonics of $\pm k^{th}$ orders display rather large intensity contrast under the periodic coding sequence 00-01-10-11. In particular, the harmonics of $+1^{st}$ and $-1^{st}$ orders have almost the same H-plane scattering pattern but make a great difference in amplitude, as shown in Fig. 3I, due to the asymmetric spectral profile. Furthermore, the modulated spectra are easily mirrored by reversing the periodic coding sequence as 11-10-01-00, as presented in Fig. 3H, since the opposite time gradient will result in the exchange of the $+k^{th}$ and $-k^{th}$ spectral lines. Such a property can be further understood by observing the scattering pattern in SOM Figure S6, in which the $\pm 1^{st}$ harmonics exchange their roles.

The large power conversion rate from carrier to the first harmonics renders the metasurface distinctive capability to realize arbitrary Doppler shift, which is especially useful for simulating echoed signals of real targets with fake Doppler frequency shift to deceive the active radar system, yielding ***velocity illusions***. For instance, a motionless time-domain digital coding metasurface with the frequency shift of $\pm$156.25 KHz (see Fig. 3G-H) may create an illusion of approaching or receding objects with extraordinary velocity $\pm$6.51 km/s to the radar operating at 3.6GHz. By seeking proper pulse width or coding sequence, dynamic Doppler shifts can be realized to mimic different moving targets with varied velocities using the programmable time-domain digital coding metasurface, which brings large challenge for conventional radar to detect the movement of camouflaged targets.

**New-architecture wireless communication system.** The time-domain coding metasurface has demonstrated excellent capability to embed the digital signals into the carrier waves at the radio frequency (RF) in free space, ***getting rid of the conventional mixing process***, which may serve as a brand new architecture for the information transmission (*36*). As illustrated in Figs.

2E-H and 3G-H, the opposite coding sequences (00-01-10-11 and 11-10-01-00) would enable efficient energy conversion from the central frequency to the $+1^{th}$ and $-1^{th}$ order harmonics, hinting that the two spectral lines can be employed as two discrete frequencies required by the traditional BFSK communication system, standing for binary information of '1' and '0' respectively inside the message.

The schematic of the new BFSK system is illustrated in Fig. 4A, in which the time-domain digital coding metasurface is used to replace the analog-digital converter (ADC) and whole RF portion composed of filters, mixers, and amplifiers in the traditional system, as shown in SOM Fig. S7. The real-time BFSK signal transmission is carried out from the meta-transmitter to a soft-defined radio (SDR) receiver (NI USRP RIO 2943R). Following the conceptual diagram in Fig. 4A, the transmission process can be divided into three steps, as presented in Fig. 4B. Firstly, FPGA generates a bit stream (such as 01101001…) of the transmitted information (e.g. pictures and movies). Then all bit streams are mapped to the corresponding coding sequences of the metasurface, which are further extended periodically to produce a pair of discrete frequencies in BFSK. Finally the electromagnetic waves containing the digital information are transmitted.

We build up the new BFSK wireless communication system, as shown in SOM Figure S8, in which the distance between the metasurface and receiver is about 6.25m. Figure S9A gives the block diagram of the receiver. All baseband algorithms for the BFSK receiver are performed on SDR platform (NI USRP RIO 2943R). The time-domain signal is transformed into the frequency domain through FFT operation and is sent to the detecting diagram to determine the spectrum intensities. The upper path in the detecting diagram is responsible for the energy detection of frequency offset $f_1$ (related to bit '1') and the lower path in the detecting diagram is responsible for the energy detection of frequency offset $f_2$ (related to bit '0'). When the power value detected in the upper path is much larger than the power value detected in the lower path, the receiver judges that the current bit transmitted by the metasurface-based transmitter is '1', and vice versa. After the detection and judgment, the bit stream is recovered and grouped by the receiver, and conveyed to a program for postprocessing. For the better illustration, SOM Figure S9B shows the instantaneously experimental result of the receiving spectrum after the FFT operation. The peak value is located at the lower frequency with the offset of -312.5KHz, which indicates that bit ‘0’ is transmitted in the current message symbol.

SOM Table S1 shows the key parameters of the BFSK wireless communication system based on the time-domain digital coding metasurface. A color picture shown in Fig. 4C is successfully transmitted over the air and recovered by the SDR receiver with the new BSFK system, and the received messages are presented in Fig. 4D-F with the receiving angles $\alpha = 0°$, $20°$ and $30°$, respectively. We clearly observe that, even though the receiver has a large alignment angle to the metasurface, the received pictures have very good quality, verifying the feasibility and superiority of the new BFSK system. More detailed descriptions on message transmission and reception are provided in SOM Video.

In summary, we have demonstrated a new theory to control nonlinearity using time-domain digital coding metasurface, whose reflection phase or amplitude can be modulated periodically with pre-defined coding sequences. This time-varying feature allows productions of high-order harmonics through the Fourier transform of the modulation signal, and hence can transfer the energy of the central frequency to harmonics of higher orders. Different from the conventional nonlinear technologies, the proposed time-domain digital coding metasurface offers substantial flexibility and accuracy in the manipulations of the harmonics by simply customizing the coding sequences in a programmable way, resulting in many unusual functions like frequency cloaking and velocity illusion. Based on the time-domain digital coding metasurface, we have proposed a novel wireless communication system with much more simplified architecture. Compared with conventional systems, the hardware complexity of the metasurface-based system is greatly simplified without degrading the system performance. The metasurface-based system also promises a significant potential to reduce the power consumption and improves the energy efficiency, and hence provides new solutions for the future wireless communication systems.

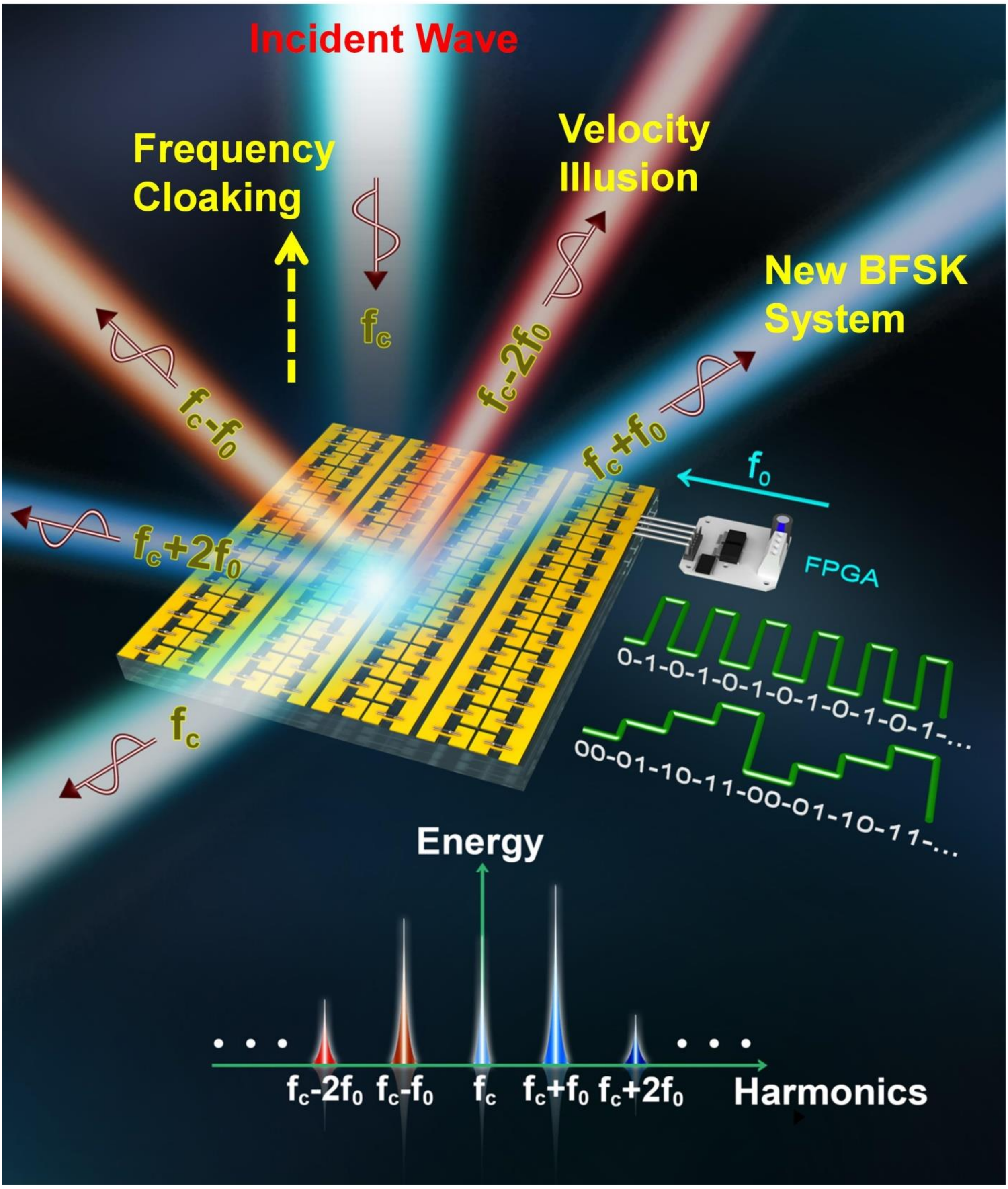

**Figure 1. Illustration of the time-domain digital coding metasurface.** The reflection phase or amplitude of meta-atoms could be electronically controlled dynamically by external biasing voltages, thus producing nonlinear modulations of spectral energies of central and harmonics frequencies by changing the coding sequence in a programmable way.

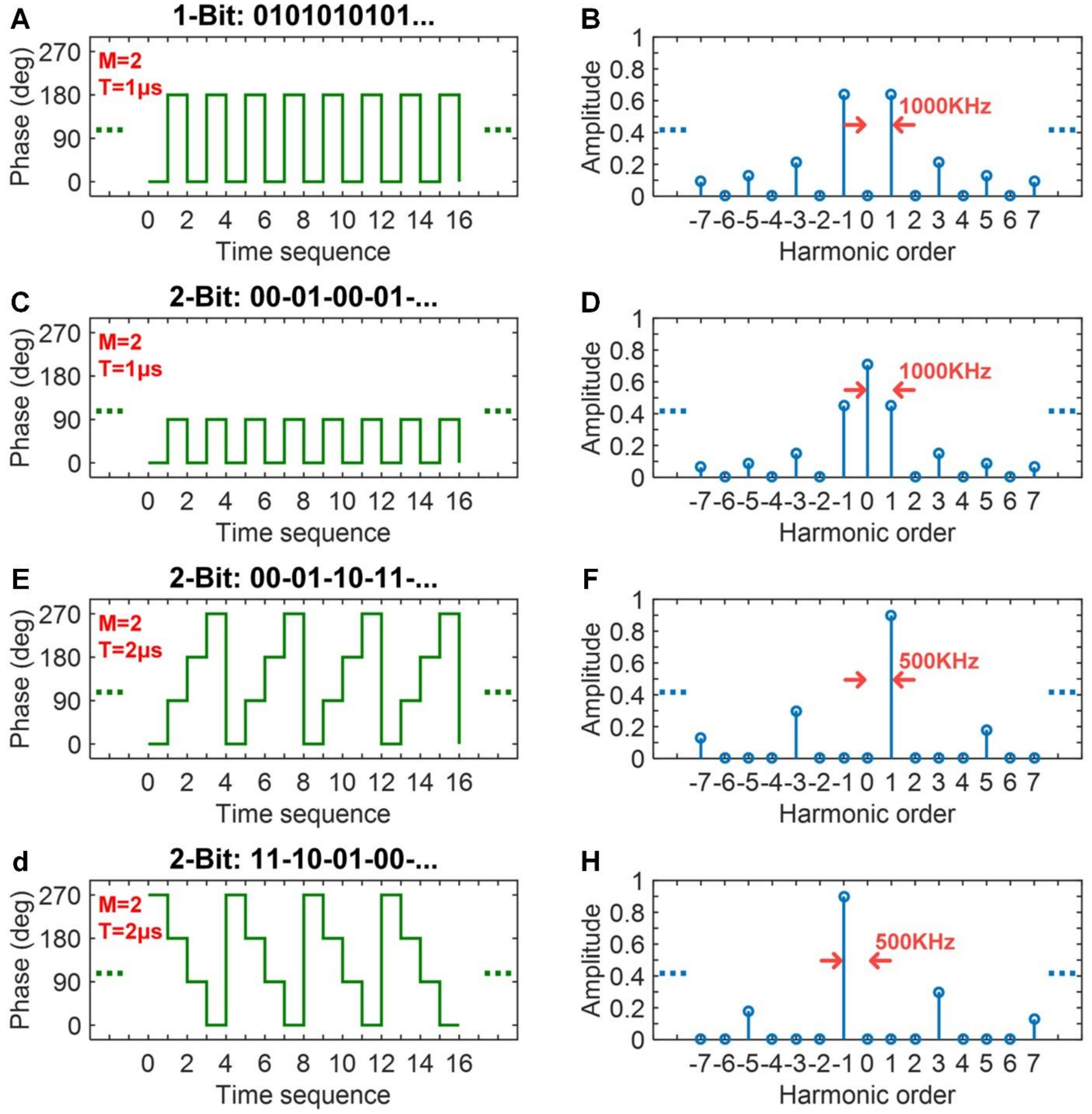

**Figure 2. The calculated spectral intensities of the output harmonics under different PM modulations. A-B,** 1-bit PM coding 01010101… with M=2 and T=1μs. **C-D,** 2-bit PM coding 00-01-00-01-… with M=2 and T=1μs. **E-F,** 2-bit PM coding 00-01-10-11-… with M=4 and T=2μs. **G-H,** 2-bit PM coding 11-10-01-00-… with M=4 and T=2μs.

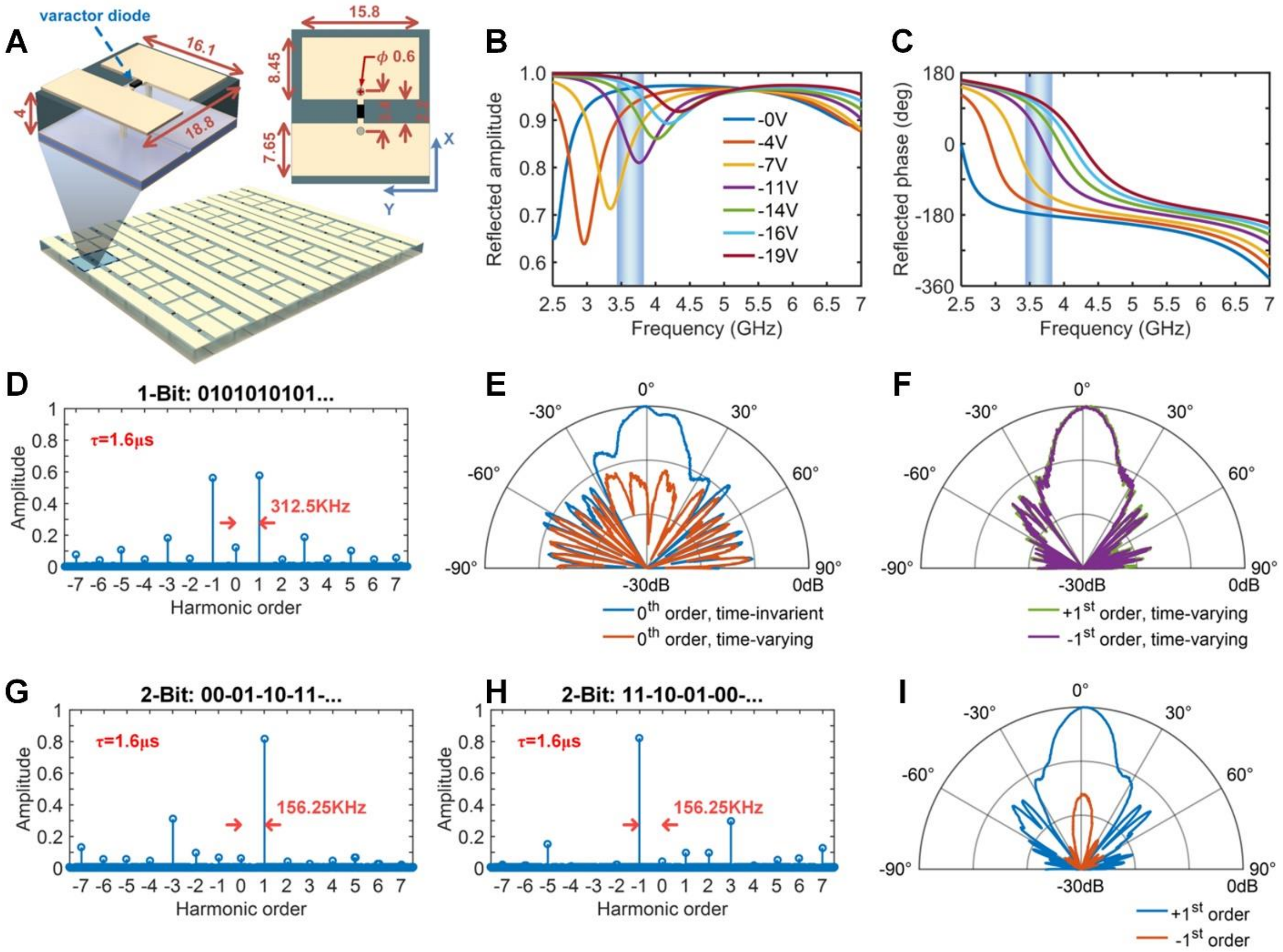

**Figure 3. A,** Schematic of the time-domain digital coding metasurface, in which the inset shows the side and front views of the meta-atom with detailed geometries. **B-C,** The simulated reflection amplitude **(B)** and phase **(C)** responses of meta-atom under different biasing voltages, in which the blue strip indicates the interested frequency region. **D,** The measured spectral intensities of all harmonics under the 1-bit coding sequence 01010101… at 3.6GHz with the pulse duration τ=1.6μs. **E,** The measured H-plane scattering patterns of fundamental harmonic modulated by the 1-bit coding sequence 01010101… (the blue line) or not (the red line). **F,** The measured H-plane scattering patterns of the +1$^{st}$ (the green line) and -1$^{st}$ (the purple line) order harmonics modulated by the 1-bit coding sequence 01010101… **G-H,** The measured spectral intensities of all harmonics under 2-bit coding sequences 00-01-10-11-… and 11-10-01-00-… at 3.6GHz, respectively. **I,** The measured H-plane scattering patterns of the +1$^{st}$ (the blue line) and -1$^{st}$ (the red line) order harmonics under the 2-bit coding sequence 00-01-10-11-…

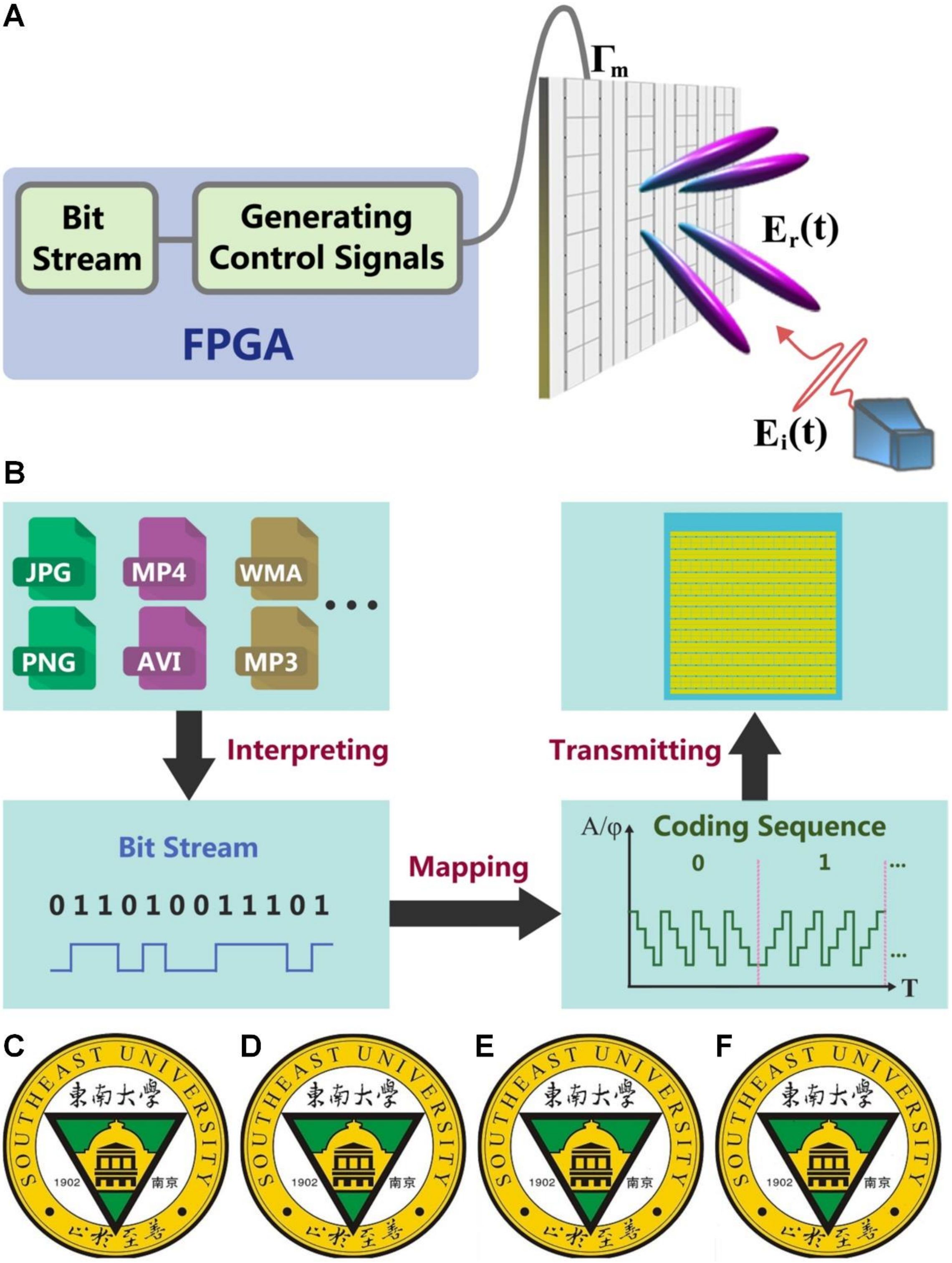

**Figure 4 A,** The schematic of the new BFSK wireless communication system based on the time-domain digital coding metasurface. **B,** The signal modulation process of the new BFSK wireless communication system. **C,** The original picture to be transmitted. **D-F,** The received messages by the new BFSK wireless communication system in experiment at different receiving angles $\alpha = 0°$, $20°$ and $30°$, respectively.

## Supporting Online Material for

# Controlling spectral energies of all harmonics in programmable way using time-domain digital coding metasurface

Jie Zhao[1†], Xi Yang[2†], Jun Yan Dai[1†], Qiang Cheng[1,3]*, Xiang Li[2], Ning Hua Qi[1], Jun Chen Ke[1], Guo Dong Bai[1], Shuo Liu[1], Shi Jin[2]*, and Tie Jun Cui[1,3]*

[1]*State Key Laboratory of Millimeter Waves, Southeast University, Nanjing 210096, China*

[2]*National Mobil Communication Research Laboratory, Southeast University, Nanjing 210096, China*

[3]*Jiangsu Cyber-Space Science & Technology Co., Ltd., 12 Mozhou East Road, Nanjing 211111, China*

† These authors contributed equally to this work.

* Corresponding author: qiangcheng@seu.edu.cn; jinshi@seu.edu.cn; tjcui@seu.edu.cn

**This PDF file includes:**

SOM Text

SOM Figure S1 to S9

SOM Equation S1 to S13

SOM Table S1

SOM Video

## 1. Derivation of Nonlinear Modulation

We take a reflective time-domain digital coding metasurface as example to elaborate the process of nonlinear modulation under monochromic incidence. As illustrated in Fig. S1, the reflectivity $\Gamma(t)$ of the metasurface is a periodic signal, and the incidence wave is defined as $E_i(t)$. Thus the reflected wave $E_r(t)$ can be expressed by the product of $\Gamma(t)$ and $E_i(t)$:

$$E_r(t) = E_i(t) \cdot \Gamma(t) \tag{S1}$$

Hence the Fourier transform can be written as

$$E_r(f) = \frac{1}{2\pi} E_i(f) * \Gamma(f) \tag{S2}$$

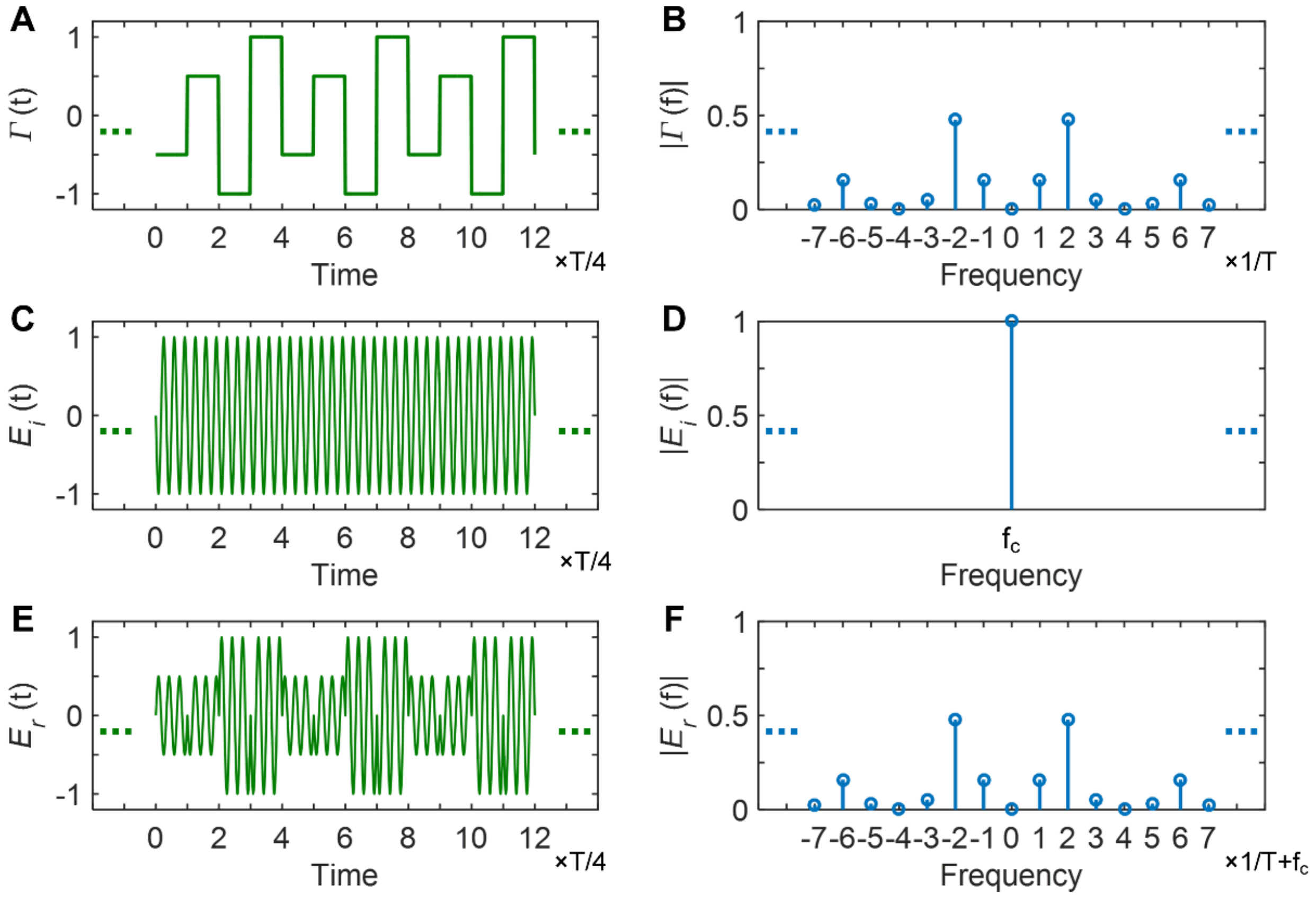

**Figure S1. An example to illustrate the nonlinear modulation in the time domain. A-B,** The reflectivity of a time-domain digital coding metasurface and its spectrum. **C-D,** The incident waveform and its spectrum. **E-F,** The time-domain reflected waveform and its spectrum.

The reflectivity function is defined over one period as a linear combination of scaled and shifted pulses

$$\Gamma(t) = \sum_{m=0}^{M-1} \Gamma_m g(t - m\tau), \qquad (0 < |t| < T) \tag{S3}$$

where $g(t)$ is the periodic pulse signal, and in each period we have

$$g(t) = \begin{cases} 1, & 0 < t < \tau \\ 0, & \text{otherwise} \end{cases} \tag{S4}$$

in which $T$ is the period of the reflectivity function and $\tau = T/M$ is the pulse width; $M$ is a positive integer greater than zero; and $\Gamma_m$ is the reflectivity at the interval $(m\text{-}1)\tau < t < m\tau$. Here the periodic pulse function $g(t)$ can be represented in alternative form of Fourier series

$$g(t) = \sum_{k=-\infty}^{\infty} c_k e^{jk2\pi t/T} \tag{S5}$$

where

$$c_k = \frac{1}{M} Sa\left(\frac{k\pi}{M}\right) \exp\left(-j\frac{k\pi}{M}\right) \tag{S6}$$

For a space-invariant system, the periodic reflectivity function can be represented as the sum of Fourier series

$$\Gamma(t) = \sum_{k=-\infty}^{\infty} a_k e^{jk2\pi f_0 t} \tag{S7}$$

and its the Fourier transform is written as

$$\Gamma(f) = 2\pi \sum_{k=-\infty}^{\infty} a_k \delta(f - kf_0) \tag{S8}$$

where $f_0 = 1/T$, and $a_k$ is the complex Fourier series coefficient at $kf_0$. Substituting Eq. (S5) and (S6) into (S3), we have

$$\begin{aligned} \Gamma(t) &= \sum_{m=0}^{M-1} \Gamma_m \cdot g(t - m\tau) = \sum_{m=0}^{M-1} \Gamma_m \cdot \left( \sum_{k=-\infty}^{\infty} c_k \exp\left(-jk2\pi \frac{m}{M}\right) e^{\frac{jk2\pi t}{T}} \right) \\ &= \sum_{k=-\infty}^{\infty} c_k \cdot \left( \sum_{m=0}^{M-1} \Gamma_m \exp\left(-j\frac{2km\pi}{M}\right) \right) e^{jk2\pi f_0 t} \end{aligned} \tag{S9}$$

Then the complex Fourier series coefficient of $\Gamma(t)$ could be expressed as:

$$a_k = \frac{1}{M} Sa\left(\frac{k\pi}{M}\right) \exp\left(-j\frac{k\pi}{M}\right) \cdot \sum_{m=0}^{M-1} \Gamma_m \exp\left(-j\frac{2km\pi}{M}\right) = UF \cdot TF \tag{S10}$$

in which,

$$TF = \sum_{m=0}^{M-1} \Gamma_m \exp\left(-j\frac{2km\pi}{M}\right), \qquad UF = \frac{1}{M} Sa\left(\frac{k\pi}{M}\right) \exp\left(-j\frac{k\pi}{M}\right) \tag{S11}$$

It is clear that $a_k$ can be regarded as the product of two terms: the time factor *TF* and unit factor *UF*. The former is related to the modulation signal within different time slots, while the latter is the Fourier series coefficient of the basic pulse $g(t)$ repeated with the period *T*. It is important to mention that the time-domain reflectivity results in a frequency conversion effect on the incoming wave and introduce a series of higher harmonics at $kf_0$. Under the excitation of a monochromatic signal $E_i(f)$ with the frequency of $f_c$, the output signal can be expressed as

$$E_r(f) = \sum_{k=-\infty}^{\infty} a_k \, E_i(f - kf_0) \tag{S12}$$

which can be further rewritten as

$$E_r(f) = a_0 E_i(f) + \sum_{k=1}^{\infty} \left[a_k \, E_i(f - kf_0) + a_{-k} E_i(f + kf_0)\right] \tag{S13}$$

## 2. Amplitude Modulation

Amplitude modulation (AM) of the reflectivity on metasurface could achieve efficient controls to the nonlinearity. Figure S2 presents the results of AM with the metasurface. We observe that the modulation function has strong impacts on the frequency offset of nonlinear components $kf_0$ and the envelope of Fourier series coefficients (or the spectrum intensities). When the surface reflectivity changes between total reflection and total absorption periodically, even-order harmonics cannot be produced except for $k = 0$, as shown in Fig. S2A-B. In contrast, for the modulated reflectivity described in Fig. S2G-H, significant suppressions of high-order harmonics are obtained, which is especially beneficial to improve the generation efficiency of the second harmonic for nonlinear applications.

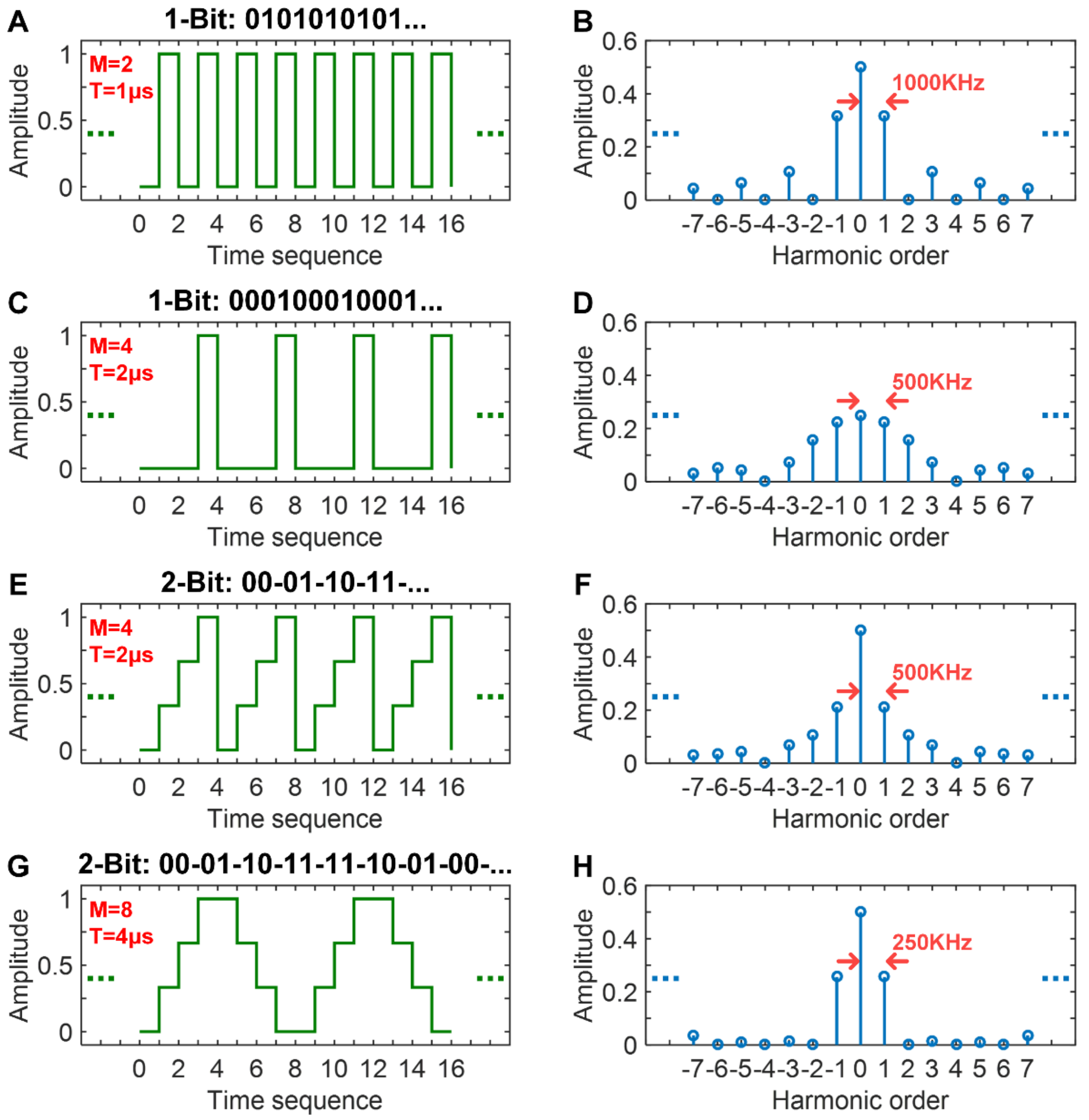

**Figure S2. The calculated spectral intensities of the output harmonics under AM. A-B,** 1-bit AM coding 01010101… with M=2 and T=1μs. **C-D,** 1-bit AM coding 00010001… with M=4 and T=2μs. **E-F,** 2-bit AM coding 00-01-10-11-… with M=4 and T=2μs. **G-H,** 2-bit AM coding 00-01-10-11-11-10-01-00-… with M=8 and T=4μs.

## 3. Experimental Setup of the Time-Domain Digital Coding Metasurface

Photograph of the fabricated time-domain digital coding metasurface sample in experimental environment is illustrated in Fig. S3A, with the schematic of experimental setup shown in Fig. S3B. In measurements, a horn antenna is used as the excitation at a distance of 1.5m, and connected to a microwave signal generator (Keysight E8257D) in microwave anechoic chamber. The other antenna is used to receive the scattering signals at a distance of 9m, and connected to a spectrum analyzer (Keysight E4447A). The operation frequency of the transmitting antenna is 3.6GHz. An interface circuit board is designed to provide the dynamic biasing voltages under the programmable control of embedded system (NI Compact RIO platform).

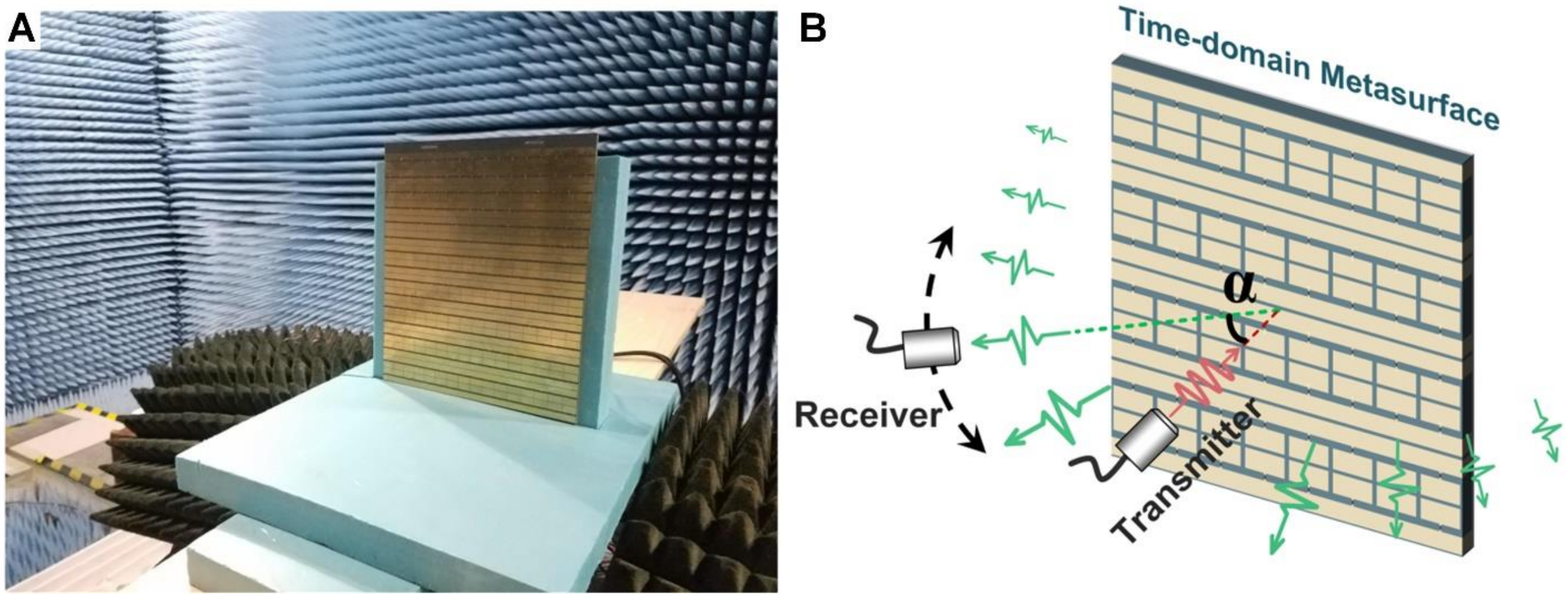

**Figure S3. A,** Photograph of the fabricated time-domain digital coding metasurface sample in experimental environment. **B,** Experimental setup to measure the spectral intensities and scattering patterns of the time-domain digital coding metasurface, where the transmitter generates electromagnetic waves in the normal direction, and the receiver is rotated along a semi-circular trajectory to monitor the field intensities.

## 4. Measured Reflections and Scattering Patterns

The measured reflection spectra under the 1-bit digital coding sequence 01010101… at 3.6GHz with different pulse durations are illustrated in Fig. S4.

We also measured the reflection spectra under different incident frequencies. Owing to the dispersion property from the meta-atom (as shown in Fig. 3C), the phase difference between the state ‘0’ (0 V) and ‘1’ (-9 V) would deviate from 180° when it is operated far from the central frequency 3.6GHz, leading to the deteriorated performance in nonlinearity generation. To find out the frequency dependence on the harmonic intensity, the time-domain coding metasurface is measured over the frequency range from 2GHz to 7GHz under the binary coding sequence 01010101…, and the reflection spectra are presented in Fig. S5A-D. Surprisingly, the metasurface appears to operate in a wide spectrum with distinct nonlinear phenomena. Theoretically the harmonic generation always exists under the modulation of periodic PM signals, even when the phase difference $\Delta\phi$ of two states deviates significantly from 180°. Figure S5E-F illustrates the spectral intensities for the $0^{th}$-order and $\pm 1^{st}$-order harmonics as the function of $\Delta\phi$. The suppression performance of the $0^{th}$-order harmonic starts to deteriorate rapidly due to the insufficient phase difference, which could be observed in the measured reflection spectra at 2, 5, and 7GHz in Fig. S5A, C, and D. We observe that the intensity of the $0^{th}$-order harmonic is much larger than that at 3.6GHz.

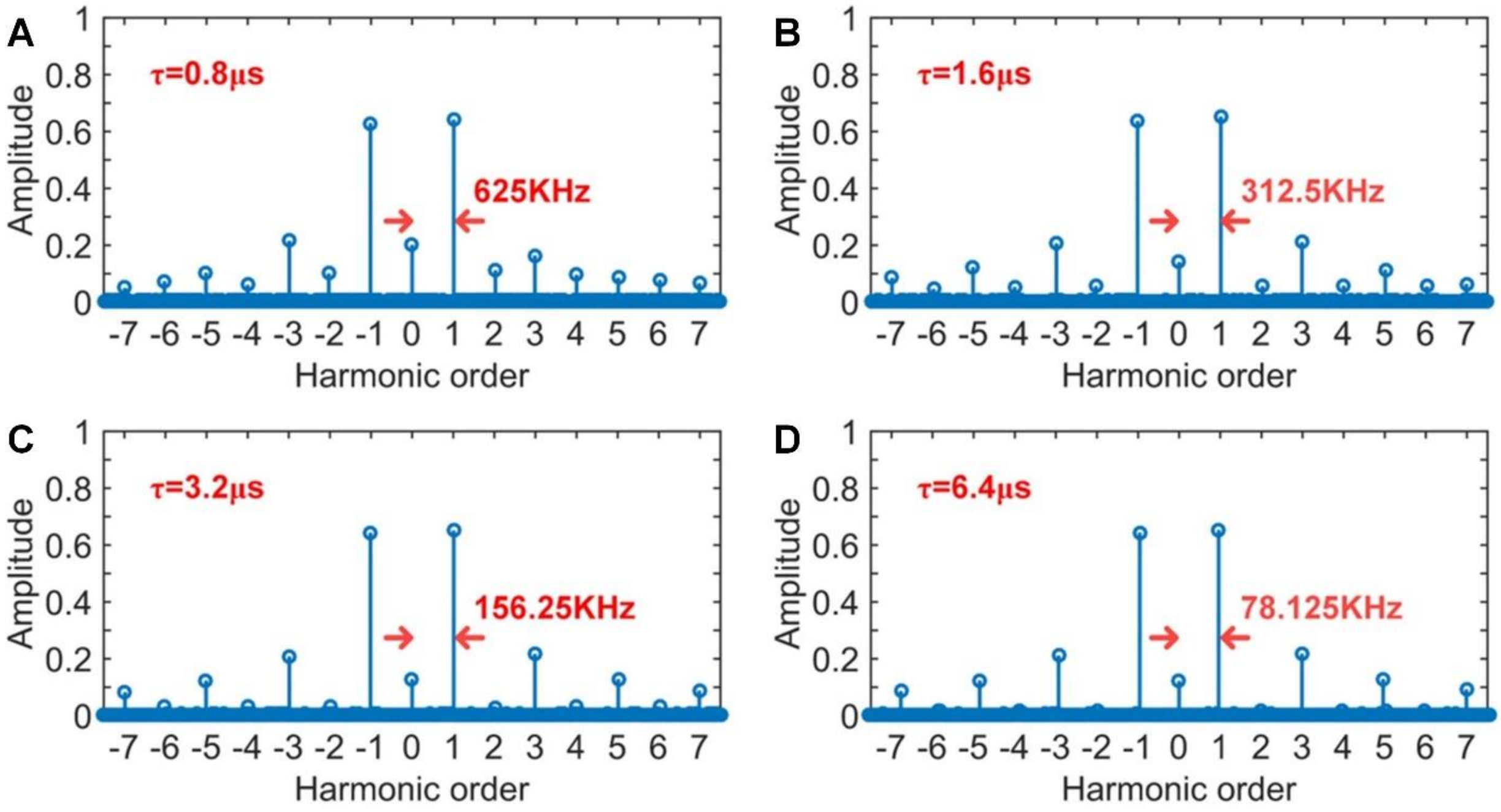

**Figure S4. The measured spectral intensities of harmonics under the 1-bit coding sequence 01010101… at 3.6GHz with different pulse durations τ. A,** τ=0.8μs. **B,** τ=1.6μs. **C,** τ=3.2μs. **D,** τ=6.4μs.

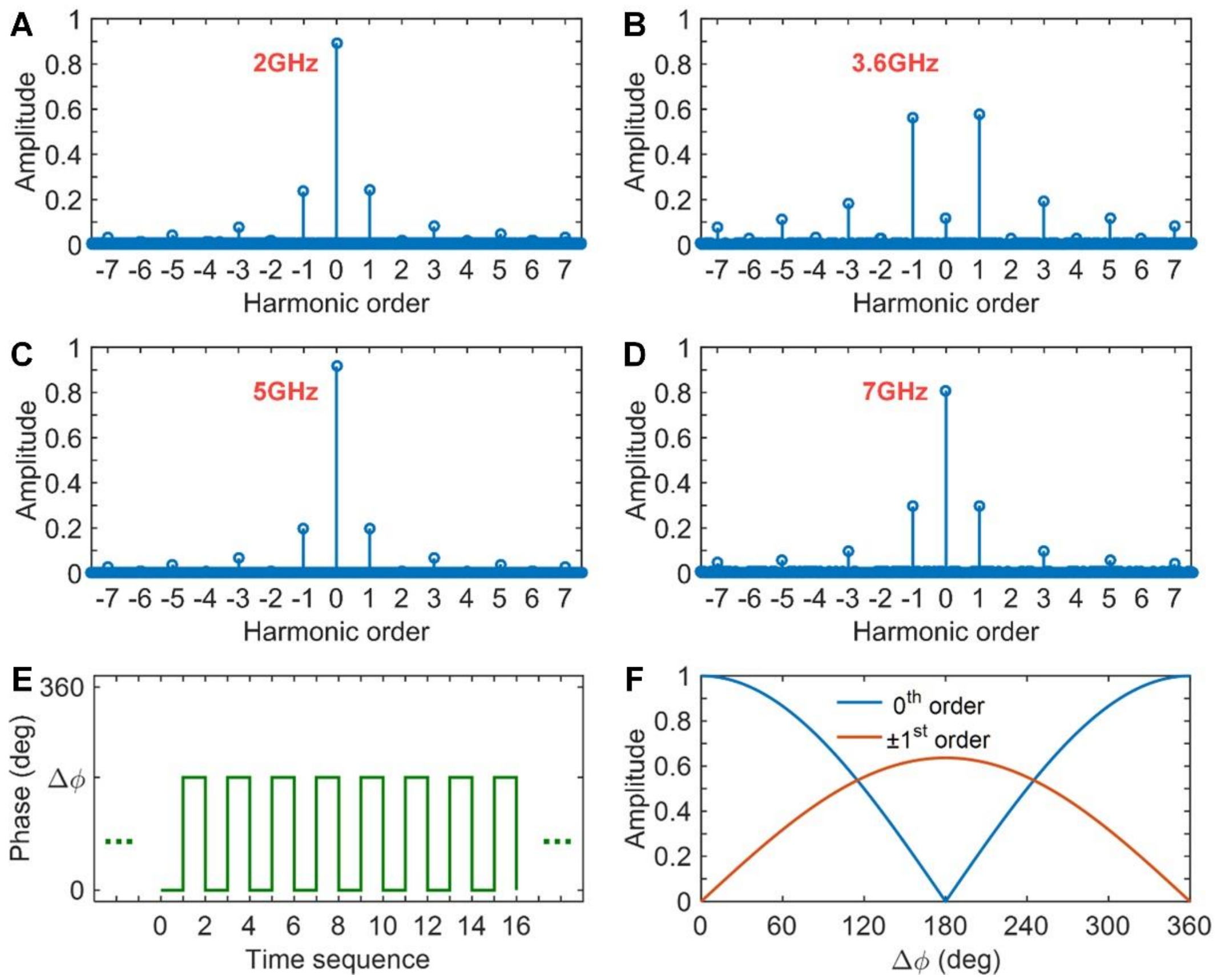

**Figure S5. A-D,** The measured spectral intensities of all harmonics under the periodic coding sequence 01010101… at 2GHz **A**, 3.6GHz **B**, 5GHz **C**, and 7GHz **D**, respectively. **E,** The phase modulation which switches between 0 and $\Delta\phi$ periodically. **F,** The theoretical spectral intensities for the $0^{th}$-order and $\pm1^{st}$-order harmonics as a function of $\Delta\phi$.

Besides the reflection spectra intensities, we have also measured the scattering patterns of the $\pm1^{st}$-order harmonics on the H-plane under the periodic coding sequence 11-10-01-00-…, as illustrated in Fig. S6.

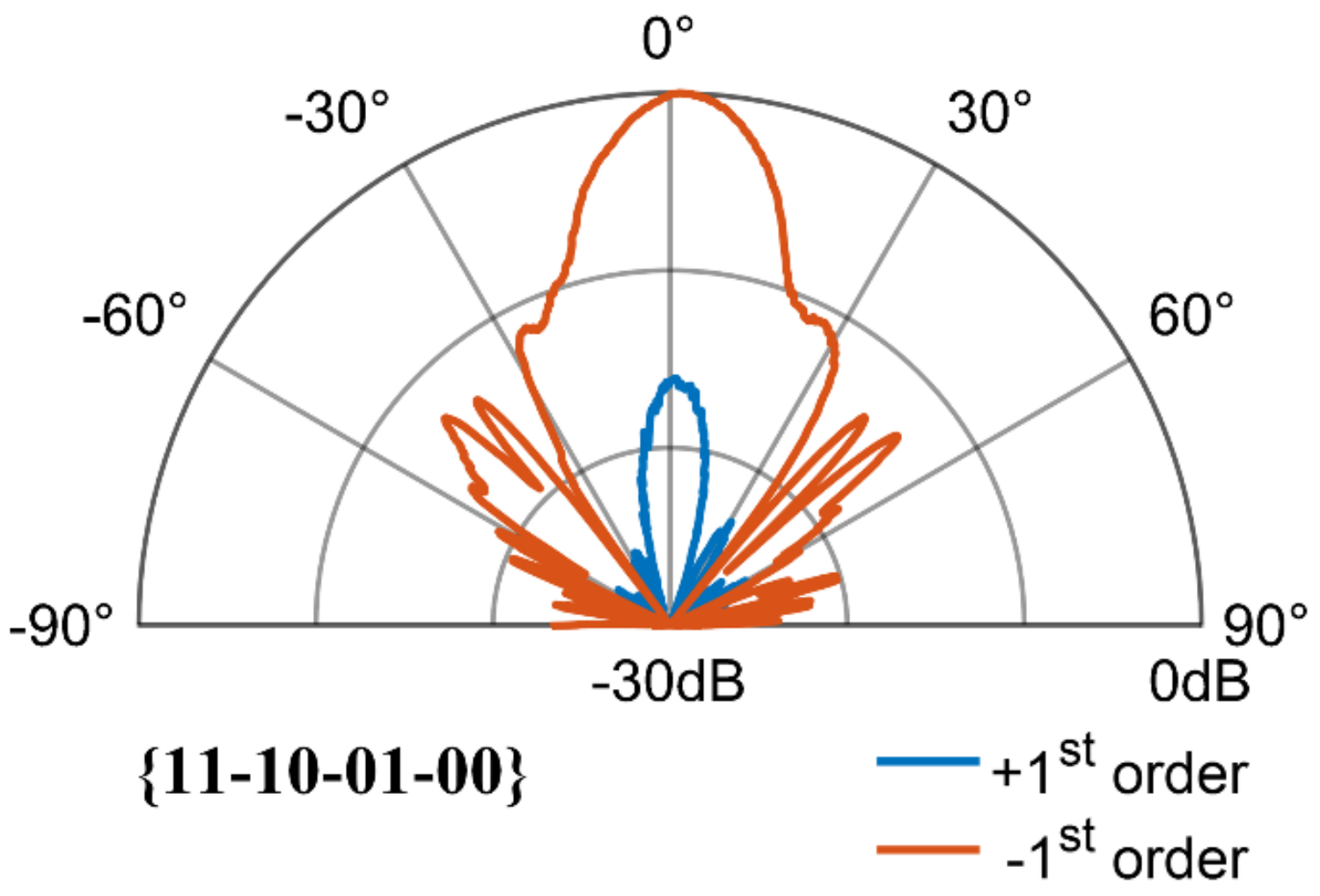

**Figure S6.** The measured H-plane scattering patterns of the +1st (the blue line) and -1st (the red line) order harmonics under the periodic coding sequence of 11-10-01-00-….

## 5. The Necessity of Digital Modulation Instead of Analog Modulation

Analog amplitude or phase modulation could actually enhance the spectral purity with excellent suppression of spurs in the echo signal from the metasurface. However, in practice the digital modulation is more advantageous due to the following reasons. Firstly, the digital modulation can greatly simply the control circuit of the metasurface since only finite phase/amplitude states are required during the modulation, and thus becomes more suitable for high-speed applications. Secondly, the analog phase modulation usually needs full-phase range to increase the nonlinear conversion rate, which is hard to achieve for the proposed metasurface by tuning the varactor at the presence of material loss. However, discrete phase states of the digital modulation are much easier to accomplish in reality, and therefore lower the barrier of element design.

## 6. Traditional Wireless Communication System

Figure S7 illustrates a schematic of the conventional super-heterodyne wireless communication system, whose architecture makes great difference from the metasurface-based system in the RF portion described in Figure 4A. The conventional wireless communication system usually consists of a digital baseband and analog RF modules, which are connected by the digital-to-analog converter (DAC). The baseband module serves as the digital signal processing center and is responsible for the implementation of various baseband signal processing algorithms; while the RF module is responsible for the up/down-conversion, carrier signal modulation, and power boosting, and hence contains a series of passive and active microwave components, including mixers, filters, low noise amplifiers, and power amplifiers, etc. The main task of the RF module is the manipulation of the carrier signal based on the message to be transmitted.

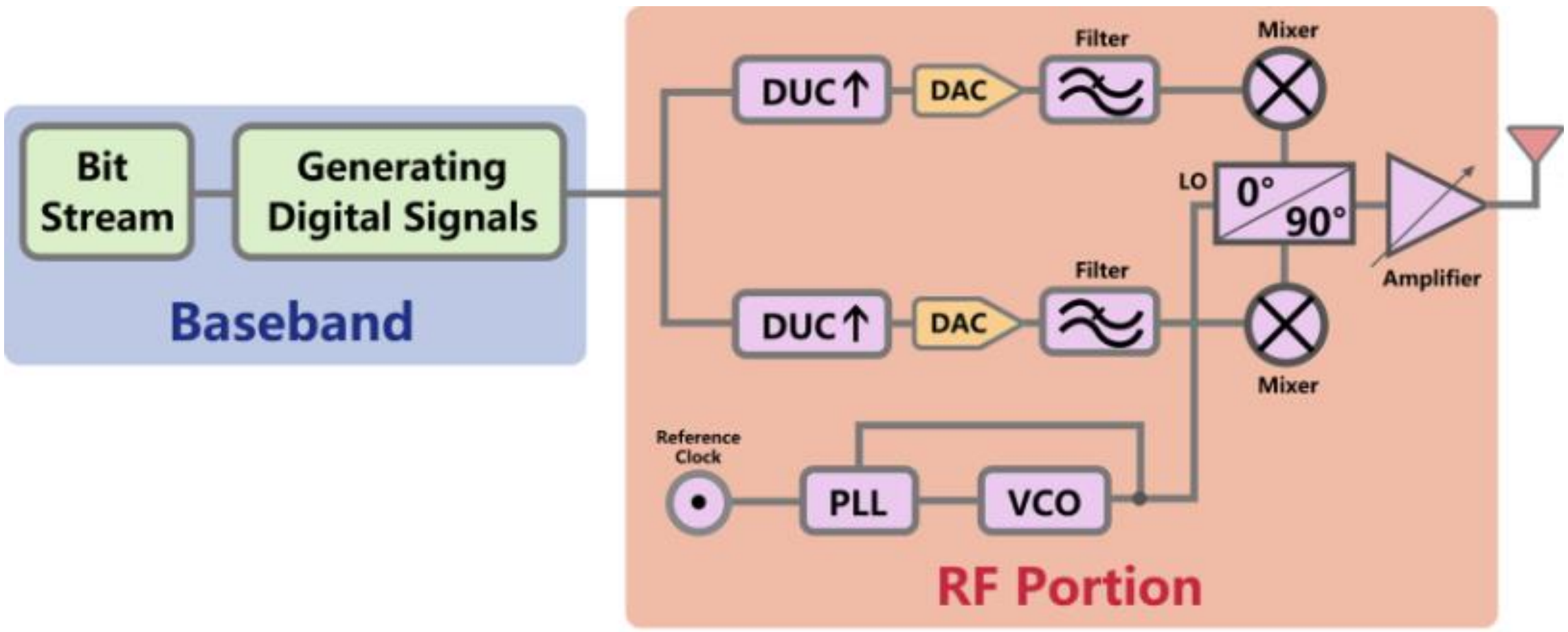

**Figure S7. The schematic of the conventional super-heterodyne wireless communication system (transmitter).**

This function can be implemented by the proposed time-domain digital coding metasurface, which can directly manipulate the incident electromagnetic waves in free space, and embed the message into the reflected waves with various coding sequences. Here we take a BFSK system for example. The message could be encoded as the bit steam like '01101001…' with the bit '1' and '0' represented by the +1st and -1st order harmonic frequencies respectively by adopting opposite coding sequences 00-01-10-11 and 11-10-01-00, as shown in Fig. 3G and 3H. Therefore, the baseband message is embedded into the reflected waves and could be handled by the receiving system.

## 7. Experimental Setup of the New BFSK Wireless Communication System

The photograph of the experimental setup for the new BSFK wireless communication system is illustrated in Fig. S8.

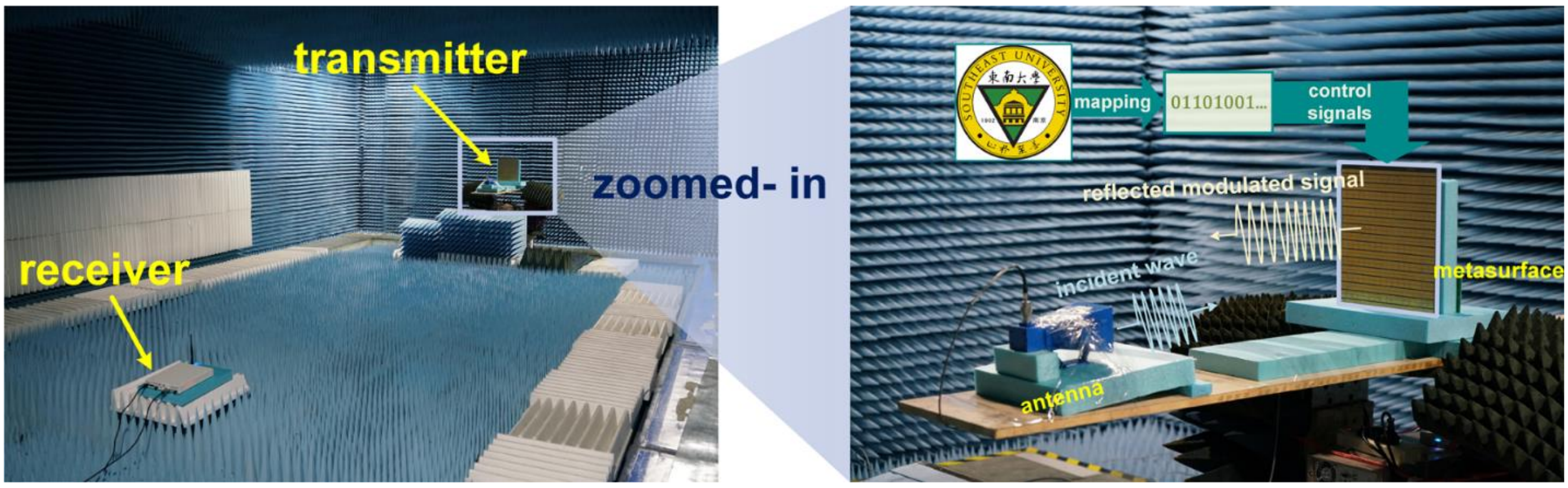

**Figure S8. The photograph of the experimental setup for the new BFSK wireless communication system.**

## 8. BFSK Receiver and System Parameters

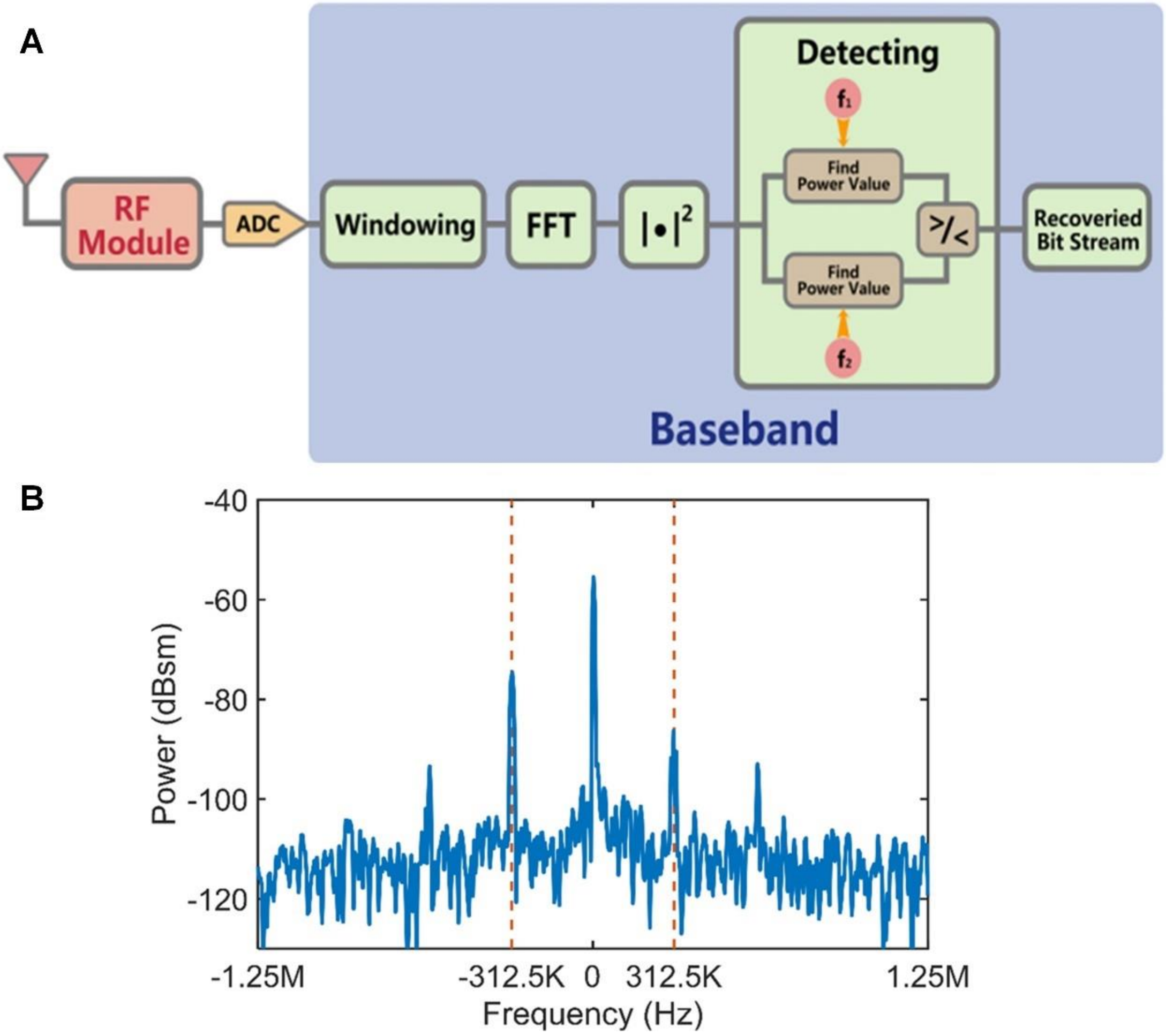

**Figure S9. A,** The block diagram of BFSK receiver. **B,** The instantaneously experimental results for the receiving spectrum after the FFT operation.

**Table S1. The primary parameters of the BFSK wireless communication system based on the time-domain digital coding metasurface**

| Parameters | Values |
|---|---|
| **Carrier frequency** | 3.6 GHz |
| **Sampling rate** | 40 MS/s |
| **FFT size** | 512 |
| **Frequency offset 1 (bit ‘1’)** | +312.5 KHz |
| **Frequency offset 2 (bit ‘0’)** | -312.5 KHz |
| **Message symbol duration** | 12.8 μs |
| **Bit rate of transmission** | 78.125 Kbps |

## 9. Video

The processes of picture transmission through the new BFSK system based on the time-domain digital coding metasurface are recorded in SOM Video, with the experimental setup depicted at 00:36 in the Video. The message, a color picture of the logo for Southeast University, is transmitted through the new BFSK transmitter, which is received by an SDR receiver placed 6.25 m away from the BFSK transmitter.

The first experiment starts at 01:01 in the Video, and the picture transmitted by the new BFSK transmitter is received and successfully recovered by receiver, validating the performance and feasibility of our system. To verify the robustness of the system, another experiment is carried out starting at 01:12 in the Video, where the new BFSK transmitter is swinging continuously between -30 degrees and 30 degrees during the message transmission. Under this circumstance, the message is received and recovered by the receiver perfectly, indicating that the new BFSK system could cover the region about at least 60° azimuth.

The last experiment is conducted to prove the characteristic of anti-interference property of our system, which could be seen in the Video starting at 01:26. An interference source is placed close to the metasurface, generating the interference signal with very high power intensity. The operation frequencies for the BFSK system are 3.6GHz±312.5KHz. In the experiment, the frequency for interference signal changes step by step from 3.6GHz+10MHz, 3.6GHz+2MHz, to 3.6GHz+550KHz, gradually approaching the operation frequency (3.6GHz+312.5KHz) of the new BFSK system. The experiment result shows that the transmitted signal from the BFSK system could still be received and recovered.

In conclusion, the proposed new BFSK system could work properly in some critical environments, and might find applications in the wireless transmission system like base station system.